\documentclass{www2012-accepted-natbibcompat}\def\newblock{\relax} 
\pdfoutput=1
\usepackage[usenames,dvipsnames]{color}
\usepackage{amsmath}
\usepackage{xspace,subfigure,url}
\usepackage{xspace}
\usepackage[numbers,sort&compress]{natbib}
\usepackage{multirow}

\usepackage{times}
\usepackage{latexsym}
\usepackage{xspace}
\usepackage{graphicx}
\usepackage{pdfsync}
\usepackage{subfigure}
\usepackage{booktabs}

\usepackage{balance}
\usepackage[dvipsnames]{color}
\usepackage[overload]{textcase}
\usepackage{amssymb}
\usepackage{centernot}
\usepackage{enumitem}

\newcommand{\xhdr}[1]{\paragraph*{{\bf #1}}}

\begin{document}
\clubpenalty=10000 
\widowpenalty = 10000

\title{
Echoes of Power:
Language Effects and Power Differences in Social Interaction
}

\numberofauthors{1}
\author{
    \alignauthor \hspace*{-1.5cm} 
        Cristian Danescu-Niculescu-Mizil \hspace{0.5cm}
        Lillian Lee \hspace{0.5cm}
        Bo Pang \hspace{0.7cm}
        Jon Kleinberg\\
    \affaddr{
        \hspace{0.8cm} Cornell University\hspace{1.5cm}
        Cornell University \hspace{0.4cm}
        Yahoo! \hspace{1.0cm} 
        Cornell University 
        } 
\newcommand{\es}{\hspace*{.2in}}
\email{\hspace*{-.2in}\large\mbox{\affaddr{cristian@cs.cornell.edu}, \es \affaddr{llee@cs.cornell.edu}, 
\es \affaddr{bopang@yahoo-inc.com}, 
     \es  \affaddr{kleinber@cs.cornell.edu}}
}
}

\maketitle

\begin{abstract}


Understanding social interaction within groups
is key to analyzing online communities.
Most 
current  work 
focuses on 
structural properties:
who talks to whom, 
and how 
such
 interactions form larger network structures.  
The interactions themselves, however, generally take place in the form
of natural language --- either spoken or written --- and one could 
reasonably suppose that 
signals 
manifested in language
 might also provide 
information about roles, status, and other aspects of the group's dynamics.  
To date, however,  finding
domain-independent language-based signals has been a challenge.

Here, we show that in group discussions, 
power
differentials between participants 
are
subtly revealed by 
how much
one individual
immediately 
echoes
the linguistic style
of the person they are
responding to.
Starting from this observation, we propose an analysis framework 
based on linguistic coordination
that can be used to
shed light on
power relationships
and that
works consistently 
across 
multiple types of 
power --- including a more ``static''
form of power based on status differences, and a more ``situational'' form
of power in which one individual experiences a type of dependence on
another.
Using this framework, we study how conversational behavior can reveal
power relationships in two very different
settings:
discussions among Wikipedians  and arguments before the
U.~S. Supreme Court.

\end{abstract}

\vspace{1mm}
\noindent
{\bf Categories and Subject Descriptors:}
J.4 [{\bf Computer Applications}]: {Social and behavioral sciences}

\vspace{1mm}
\noindent
{\bf General Terms:} Measurement, Experimentation

\vspace{1mm}


\newcommand{\vpara}[1]{\vspace{0.05in}\noindent\textbf{#1. }}

\newcommand{\powerhyp}{\ensuremath{{\mathcal{P}}}\xspace}
\newcommand{\fromtwo}{\ensuremath{{\mathcal{B}}}\xspace}
\newcommand{\toone}{\ensuremath{{\mathcal{P}_{target}}}\xspace}
\newcommand{\fromone}{\ensuremath{{\mathcal{P}_{speaker}}}\xspace}
\newcommand{\fromoneprime}{\ensuremath{{\mathcal{P}_{speaker}'}}\xspace}

\newcommand{\court}{Supreme Court\xspace}

\newcommand{\statuseffect}{{{\color{Green}{[Status]}}}\xspace}
\newcommand{\baselineeffect}{{{\color{Orange}{[Baseline]}}}\xspace}
\newcommand{\reinforcementeffect}{\statuseffect}
\newcommand{\needeffect}{\statuseffect}

\newcommand{\wikidatashort}{{\it wiki}\xspace}
\newcommand{\courtdatarestricted}{{\it court}\xspace}
\newcommand{\coordfeature}{coordination features\xspace}
\newcommand{\baselinefeature}{stylistic features\xspace}
\newcommand{\bowfeature}{bag of words\xspace}
\newcommand{\samemean}{samemean\xspace}
\newcommand{\reply}[1]{\ensuremath{R_{#1}\xspace}}
\newcommand{\length}[1]{\ensuremath{Len(#1)\xspace}}
\newcommand{\preda}{\ensuremath{x}\xspace}
\newcommand{\predb}{\ensuremath{y}\xspace}
\newcommand{\replya}{\reply{\preda}\xspace}
\newcommand{\replyb}{\reply{\predb}\xspace}
\newcommand{\sfmath}{\ensuremath{\mathcal{F}_{style}}\xspace}
\newcommand{\indomain}{in-domain\xspace}
\newcommand{\crossdomain}{cross-domain\xspace}
\newcommand{\consequential}{consequential conversations\xspace}

\newcommand{\markhigh}[1]{{\color{Purple}#1}}
\newcommand{\marklow}[1]{{\color{Green}#1}}

\newcommand{\high}{\markhigh{\em high-powered}\xspace}
\newcommand{\higher}{{\em higher-powered}\xspace}
\newcommand{\low}{\marklow{\em low-powered}\xspace}
\newcommand{\High}{\markhigh{\em High-powered}\xspace}
\newcommand{\Low}{\marklow{\em Low-powered}\xspace}

\newcommand{\adminstobe}{\marklow{admins-to-be}\xspace}
\newcommand{\failedtobe}{\marklow{failed-to-be}\xspace}
\newcommand{\failedtobehigh}{\markhigh{failed-to-be}\xspace}

\newcommand{\Section}{\S}
\newcommand{\admins}{\markhigh{admins}\xspace}
\newcommand{\nonadmins}{\marklow{non-admins}\xspace}

\newcommand{\justice}{\markhigh{Justice}\xspace}
\newcommand{\justices}{\markhigh{Justices}\xspace}
\newcommand{\lawyer}{\marklow{lawyer}\xspace}
\newcommand{\lawyers}{\marklow{lawyers}\xspace}

\newcommand{\dependence}{dependence\xspace}
\newcommand{\defendent}{dependence\xspace}
\newcommand{\experienced}{\markhigh{experienced lawyers}\xspace}
\newcommand{\novice}{\marklow{novice lawyers}\xspace}

\newcommand{\marker}{marker\xspace}
\newcommand{\markers}{markers\xspace}

\newcommand{\unfavorable}{\markhigh{unfavorable Justice}\xspace}
\newcommand{\favorable}{\marklow{favorable Justice}\xspace}
\newcommand{\unfavorables}{\markhigh{unfavorable Justices}\xspace}
\newcommand{\favorables}{\marklow{favorable Justices}\xspace}
\newcommand{\unfavorableshort}{\markhigh{unfav. Justice}\xspace}
\newcommand{\favorableshort}{\marklow{fav. Justice}\xspace}

\newcommand{\different}{\markhigh{diff. vote}\xspace}
\newcommand{\same}{\marklow{same vote}\xspace}

\newcommand{\higherc}{\markhigh{\textbf{higher power}}}
\newcommand{\lowerc}{\marklow{\textbf{lower power}}}

\newcommand{\speaker}{speaker\xspace}
\newcommand{\target}{target\xspace}
\newcommand{\speakers}{speakers\xspace}
\newcommand{\targets}{targets\xspace}


\section{Introduction}
\label{sec:intro}

With the arrival of detailed data on the interactions within social
groups --- generally coming from the on-line domain --- 
an active line of research has developed around the phenomena 
taking place in these groups.
To date, these analyses have mainly used structural features of
the interactions, including who talks to whom, how frequently, 
and how these patterns of interaction form larger network structures.

But the interactions themselves are generally taking place in natural 
language --- both spoken and written --- and the language content of
these interactions has been a long-acknowledged missing ingredient in
this style of investigation.
The reason for this is clear: while it is reasonable to suppose 
that signals within the language could provide insight into 
the social structure of the group, it has been challenging
to extract useful language-level signals.
A small but growing line of work has begun to use textual content
for uncovering structural properties of on-line networks
\cite{hub2004,Diehl+Namata+Getoor:2007a,Gilbert2009,Choudhury2010,Menczer,Bramsen+al:2011a,livne2011party,Gilbert:ProceedingsOfCscw:2012,Otterbacher:ProceedingsOfCscw:2012};
it is exciting to contemplate extending
the range of social properties
that can be analyzed via text.

\xhdr{Power and linguistic style}
In this paper, we show how variations in linguistic style can
provide information about power differences 
within social groups.
Our focus is on domains in which groups engage in goal-oriented
discussions --- situations where people interact, 
not necessarily collaboratively, in order to accomplish tasks
or settle on choices.
An important
characteristic of
such discussions
is that the participants are
invested in the issues at hand, so that their dialogs are not simply
``idle chat'', but consequential: the outcome matters.
Examples include conversations among
wiki editors or open-source teams regarding modifications;
debates within conference program committees on which papers to accept;
and discussions in legal hearings, where opposing sides compete
to persuade a judge or jury.

Power differences among the participants constitute a crucial force in all
these settings.  Sometimes these power differences are embodied in formal
roles, such as that of a judge or a program chair.
Sometimes they are based on more informal differences in the respect
or authority commanded by individuals within the group.
And sometimes they are more situational: \preda may have power over \predb
in a given situation because \predb needs something that \preda can choose
to provide or not.

It is natural to ask how we might try 
to create widely-applicable methods for
inferring these power
differences
simply by observation of the language used 
within a group.
This is particularly challenging if we are seeking methods that
generalize across domains, and are not tied to specific choices of content.
By way of analogy, imagine that you walk into a meeting among people
you've never met, and on a topic that you know nothing about;
what could you do to identify who are the most powerful members of the group?
If you were actually able to observe the people and hear them
speaking to each other, then
cues such as posture and vocal pitch can provide such information 
\citep{Giles:EngagingTheoriesInInterpersonalCommunicationMultiplePerspectives:2008,huang-posture}.
But if we only have the 
text or transcripts of their interactions --- the formats that
online
data often takes --- 
how do we identify evidence of power differences? 

\xhdr{Language coordination}
We propose that
{\em language coordination} in text content alone
can serve as a rich source of information about power differences
within a group.
Language coordination is a phenomenon in which 
people tend to unconsciously mimic the choices of function-word
classes made by the people they are communicating with \cite{Niederhoffer+Pennebaker:2002a};
roughly speaking, if you are communicating with someone who uses
a lot of articles --- or prepositions, or personal pronouns --- 
then you will tend to increase your usage of these types of words as well,
even if you don't consciously realize
it.\footnote{We note that language coordination is just one form of coordination
where such phenomena occur \cite{giles19911} (another is posture coordination, for example);
we focus on language coordination because it can be measured in
textual 
interactions.}

We measure language coordination in two datasets of goal-oriented 
text that arise in very different settings: 
discussions among Wikipedia editors, containing
over 240,000 conversational exchanges; and oral arguments before
the U.S. Supreme Court, as processed by Hawes et al. 
\citep{Hawes2009,Hawes:JournalOfTheAmericanSocietyForInformation:2009}
and containing 50,389 conversational exchanges among
Justices and lawyers.
By focusing on function word classes,
rather than domain-specific substantive content, we are able
to evaluate the domain-independence of our techniques and their 
ability to generalize across different contexts;
methods that rely on subject-specific cues to determine levels of
power (such as the use of ``Your honor'' in a legal setting) are
not positioned to generalize as readily.

To be able to speak in a principled way about power differences,
we draw on the framework of {\em exchange theory} from sociology
\citep{willer-exchange-book}.
Exchange theory and its generalizations \citep{thye-exchange-status}
have distinguished between two forms of power, which naturally parallel
the types of power in our discussion above.
First, a power difference between \preda and \predb can be based on the
fact that \preda has higher {\em status} than \predb, either through
a formal designation of status, or through more informal notions
of status based on reputation within the group.
Second, a power difference can also arise through {\em dependence}: 
if \predb needs something from \preda, and hence is dependent on \preda,
this can give \preda a form of 
at least temporary power over \predb.

\xhdr{Power differences from language coordination}
We find that differences in the level of language coordination
consistently reveal
both of these types of power differences, in both of our datasets.
Specifically, we will present the following results.
\begin{enumerate}
\item In general, people with low power exhibit greater language
coordination 
than people with high power.
\item Conversely, people coordinate more with 
interlocutors
 who
have higher power than with those who have lower power.
\item When a person undergoes a change in status, their coordination
behavior changes, and so does the coordination
behavior of people talking to them.
\item When an individual is trying to convince someone who holds
an opposing view, this creates a form of dependence and hence
a power deficit in the sense of exchange theory;
we find increased levels of language coordination in such cases.
\item The relation between status level and the extent of language coordination
transfers across domains, and  is a reliable cross-domain feature for
status prediction.
\end{enumerate}

These results suggest clear potential applications to the analysis
of on-line social groups.
In particular, they could provide methods for identifying power differences
and levels of status in on-line settings where one has only the text
content of social interactions, rather than explicit markers of 
status or explicitly annotated links.
Similarly, they could also provide a means of analyzing conversations
between
users of a social media platform so as to determine the
power balance or levels of relative status in their relationship.
In all such uses, the methods do not require domain-specific
knowledge of the on-line application being analyzed.
We also note that the role of features internal to the content
can be crucial in some of these settings, since it has been observed
that message frequency and message volume do not necessarily suffice
to determine relative status.
As Rowe et al.~state 
\citep{Rowe:2007:ASH:1348549.1348562},
``As we move down the corporate ladder, the conversational flows 
of dissimilar employees can in fact be quite similar.''
Indeed, it is easy to think of contexts where 
dominant individuals consume a lot of
the conversational bandwidth, and others where,
contrariwise,
 low-status 
individual take up most of the airtime with their advocacy
toward higher-status participants.

There is something striking about the fact that
the content
 features being employed are properties of
language that tend to escape conscious attention.
The phenomena we find in the text content are consistent and significant,
but they are not effects one notices in reading or listening
to the interactions; in essence, they operate on levels that only
show up when you use computational methods to explicitly tune
 in to them.
Moreover, since our methods are based on function words,
it means one can apply them to language
samples from which the content words have been redacted,
raising intriguing implications for compact representations
and user privacy.

\xhdr{Summary: Novel contributions of present work}
Our use of language coordination as a key source of information
draws on a history of coordination studies originating 
in social psychology;
we discuss this background in 
\S \ref{sec:hyp}.
These psychological studies of coordination focused on small-scale settings
where participant behaviors could be individually observed;
the identification of language coordination phenomena in large-scale
on-line text was done
recently by
\cite{Danescu-Niculescu-Mizil+al:11a}
using data from Twitter.
To our knowledge, our work is the first to identify
connections between language coordination and social power relations
at large scales, and across a diverse set of individuals and domains.

In addition, our work here provides the following further
novel contributions. 

{\em Multiple domains with large amounts of data.}  By using large amounts
of data, we can pick up subtle effects and explicitly vary
some of the underlying conditions for coordination across different
subsets of the data.  Moreover, 
working with two 
different corpora allows us to
test the domain independence of our linguistic-coordination
approach. 

{\em Status change.} 
Wikipedians can be promoted to administrator status 
through a public election, and almost always
after extensive prior involvement in the community.
Since we track the communications of editors over time, 
we can examine how linguistic coordination
behavior changes when a Wikipedian becomes an ``admin''.  To our
knowledge, our study is the first to analyze the effects of status
change on specific forms of
language use.

{\em Situation-dependent forms of power.}
By generalizing from status to broader notions of power,
our study is, to our knowledge, also the first to show
how multiple types of power relationships --- and in particular
situation-dependent power --- 
can be exposed through domain-independent textual features.


\section{Coordination and power}
\label{sec:hyp}

We can apply communication accommodation theory
\citep{Natale:1975,giles19911,Street:1982,Giles:EngagingTheoriesInInterpersonalCommunicationMultiplePerspectives:2008},
an influential line of research in sociolinguistics, 
to our investigations because the theory implies  the following principle:
\begin{description}
\item Principle \powerhyp.  
Linguistic coordination is a function of the power differential between the speaker and the target: the lower the  power of the speaker relative to that of the target, the more she coordinates (and vice versa, the higher the relative power of the speaker, the less she coordinates).
\end{description}
Here and throughout, {\em speaker} refers to the person producing the 
reply in an exchange,
and {\em target} refers to the person 
initiating the exchange (and thus the target of the speaker's reply).
\footnote{
We use ``initiate'' and ``reply'' loosely: in our terminology,  
the
conversation $\langle \preda$: ``Hi.'' $\predb$: ``Tired?''  $\preda$:
``No.''$\rangle$ has two exchanges, one initiated by $\preda$'s ``Hi'',
the other by $\predb$'s ``Tired?''. 
}
In the context of group conversations, which is the focus of the present work, this principle leads to the following two concrete 
hypotheses, 
based on the 
\underline{p}ower of the target and of the speaker, respectively:
\begin{description}
\item \toone: People
in general
 coordinate more towards {\em high-powered} people than towards 
{\em low-powered} 
people.

\item \fromone: {\em High-powered} people 
coordinate 
less
than {\em low-powered} people
towards their targets.
\end{description}
(Neither hypothesis implies the other because we employ an asymmetric definition of coordination.)

In addition to power imbalance,
we hypothesize that 
personal traits of the participants 
also 
influence
how much they coordinate:
\begin{description}
\item \fromtwo. 
People have a 
\underline{b}aseline coordination level, which is determined by personal characteristics
(such as their sociability and level of social engagement).
\end{description}

It is worth noting that it is not actually {\em a priori} obvious that  \toone and \fromone hold at large. First, there are 
competing  theories which postulate that the relation between power and coordination is the reverse of \powerhyp,  due to a desire of high-status individuals to be understood \citep{Bell:84a}.
Second, empirical studies supporting the hypotheses above are, while intriguing, relatively small in scale.
For example, 
\cite{Gregory:JournalOfPersonalityAndSocialPsychology:1996} showed that Larry King, the host of a popular talk-show in the U.~S., coordinated more in his vocal pitch to his high-status guests (such as then-President Clinton) than to low-status guests.
As for
{\em linguistic style} coordination, \citep{Niederhoffer+Pennebaker:2002a} looked at 15 Watergate
transcripts 
involving only four people altogether (Richard Nixon and three of his
aides); small numbers of courtroom trials have also been considered
\cite{Aronsson:JournalOfLanguageAndSocialPsychology:1987,Erickson+al:78a}.

While power might correlate with certain personal traits in a given community, making the distinction between \powerhyp and \fromtwo difficult,
they differ in one important aspect
 which we will exploit in our study:
power can change abruptly --- such as when an individual is assigned
a new role --- while personal traits, in comparison, are more stable over time.
As a result, examining the temporal change in coordination level of people
who have undergone changes in power can help us isolate the effect of 
\powerhyp
from that of \fromtwo.
In particular, this will help us address the following question:
if we do find evidence supporting hypothesis \fromtwo, would it be sufficient
to explain the data, or will we see power playing a role on top of 
baseline individual coordination levels?


\section{Power Relations in Wikipedia and Supreme Court Data}
\label{sec:data}

In this section, we describe the two corpora of consequential
discussions we used in our studies.  The first consists of discussions
between editors on Wikipedia; the second consists of transcripts of
oral arguments before the United States Supreme Court.  Both settings
involve power differentials, both through status and dependence,
as we will see below.
Our Wikipedia corpus is much larger,
potentially more representative of online discussions, and allows us
to study the effects of changes in power; but the Supreme Court
represents a less collaborative situation than Wikipedia
(in the Supreme Court data, there are always explicit opposing sides)
and is an instance of an off-line setting.  The
differences in the two corpora help us focus on general,
domain-independent relationships between relative power and linguistic
coordination.

We begin by briefly describing the roles and text content of our two domains,
and then discuss how we formalize the different kinds of power imbalances
within the domains.

We will release our data publicly at
\url{http://www.cs.cornell.edu/~cristian/www2012/}. 

\subsection{Discussions among Wikipedia editors}
\label{sec:data:wiki}

\vpara{Roles and role changes}
Wikipedia editors form a close community with salient markers of status.
Administrators, commonly known as {\em admins}, are Wikipedia editors ``trusted with access to restricted technical 
features'' such as 
protecting or deleting pages or blocking other editors\footnote{\url{http://en.wikipedia.org/wiki/Wikipedia:Administrators}}.
In effect, admins have a higher status than other users ({\em non-admins}) in the Wikipedia community,
and editors seem to be well aware of the status and activity history of other editors.
Users are promoted to admins through a transparent election process known as requests for
adminship\footnote{\url{http://en.wikipedia.org/wiki/Wikipedia:Requests_for_adminship}}, or {\em RfAs},
where the community decides who will become admins.
Since RfAs are well documented and timestamped, not only do we have the current status of editors,
we can also extract the exact time when editors underwent role changes from non-admins to admins.

\vpara{Textual exchanges}
Editors on Wikipedia interact on {\em talk} pages\footnote{\url{http://en.wikipedia.org/wiki/Wikipedia:Talk_page_guidelines}}
to discuss changes to article or project pages.
We gathered
240,436 conversational exchanges 
carried out on the talk pages,
where the participants of these 
(asynchronous) discussions were associated with rich status 
and social interaction information: status, timestamp of status change if there is one, 
and
 activity level on talk pages, which can serve as a proxy of 
editors'
sociability, or how
socially inclined they are.
In addition, 
there is a discussion phase during RfAs, where users ``give their
opinions, ask questions, and make comments'' 
about an open nomination.
Candidates can reply to existing posts during this time.
We 
extracted conversations that occurred in RfA discussions,
and obtained a total of 32,000 conversational exchanges.
Most of our experiments were carried out on the larger dataset
extracted from talk pages, unless otherwise noted.

\subsection{\court oral arguments}
\label{sec:data:court}

While Wikipedia discussions provide a large-scale dataset with rich meta-information,
overall, high-status people and low-status people are collaborating to accomplish a task.
Other social hierarchies involve much less collaboration or even
explicitly adversarial relationships.
Oral arguments before the \court provide such a setting.

\vpara{Roles}
A full court consists of  nine Justices, although
occasionally some recuse themselves.
In the oral arguments for a case, lawyers for each party have thirty minutes to present their side to the Justices. The Justices may interrupt these presentations with comments or questions, leading to interactions between the lawyers
(plus amici curiae, who for our status-based investigations count as lawyers)
 and Justices. 
After the oral arguments and subsequent deliberations, cases are decided by majority vote of the Justices.
This provides an interesting additional test ground: instead of asynchronous textual exchanges in a social hierarchy working collaboratively,
here we have verbal exchanges in a social hierarchy where {\em Justices} decide the final outcome. 
In addition, conversations here are over topics in a completely different domain.

\vpara{Transcripts of verbal exchanges}
Transcripts of oral arguments in Supreme Court are publicly available\footnote{\url{http://www.supremecourt.gov/oral_arguments/}}.
We used a pre-processed version of this dataset 
described in \cite{Hawes2009}.
We enhanced this dataset with the final votes from the Spaeth
Supreme Court database\footnote{\url{http://scdb.wustl.edu/}}.
In total, we have 50,389 verbal exchanges for 204 cases.
11 justices (two of which have little conversational data:
  Thomas\footnote{In 2011, Justice Thomas marked five terms without
    speaking in any oral arguments. \citep{Liptak:2011a}
} and Alito)
and 311 lawyers are represented in the dataset.
73\% of the lawyers only appear in one case, and the maximum number of cases 
where one lawyer appears is 15.
As such, trends identified on this dataset
should not be due to idiosyncratic behavior of a few over-represented lawyers.

\subsection{Power Relations in the Data}
\label{sec:powerindata}

\begin{table}
\begin{tabular}{r|cc}
&\multicolumn{2}{c}{Wikipedia }  \\
 &\higherc&\lowerc \\ \toprule
\multirow{2}*{Status} & \admins &\nonadmins		 \\
& \admins  & \adminstobe\ \marklow{(before RfAs)}\\ \midrule
Dependence & \different	&\same	\\ 
\bottomrule
\\
\end{tabular}
\begin{tabular}{r|cc}
&\multicolumn{2}{c}{\court} \\
 &\higherc  &\lowerc\\ \toprule

\multirow{2}*{Status} 		&\justices &\lawyers \\	
&\markhigh{Chief Justices}&\marklow{Associate Justices}\\
\midrule
Dependence	&\unfavorable &\favorable \\\bottomrule
\end{tabular}
\caption{Power differentials exhibited in our data.}
\label{tab:differentials}
\vspace*{-0.7cm}
\end{table}

Having now surveyed the nature of the two domains, we discuss
the different kinds of power relations that they contain.
An overview of the following discussion is summarized 
in Table \ref{tab:differentials}.\footnote{Throughout the paper we use
  color coding to indicate the relative power relations relevant for
  the respective discussion.  These colors are simply
 intended as a helpful
mnemonic and can be ignored without any loss of meaning.
}

In our discussion of roles earlier in this section, we have
already indicated some of the basic status differences:
the distinction between \markhigh{admins} and \marklow{non-admins} 
on Wikipedia,
and the distinction between 
\markhigh{Justices} and \marklow{lawyers}
 in the context of the Supreme Court.
We can also identify certain finer-grained distinctions, including 
the distinction between the \markhigh{Chief Justice} of the Supreme Court
(our data overlaps the terms of two 
different 
Chief Justices)
and the \marklow{Associate Justices}.
And on Wikipedia, we can also study the behavior over time of users 
who were promoted to the position of admin --- in effect, comparing
their behavior as \markhigh{admins} to their earlier behavior
as \marklow{admins-to-be}.

Our data also makes it possible to study several instances of
power differences based on {\em dependence}.
To begin with, we note the general principle that 
status and dependence are almost never completely distinct
\cite{thye-status-value},
since a person in a high-status role frequently appears in situations
where people are dependent on them.

The data, however, offers us opportunities to study forms of dependence
where the level of status has been largely controlled for.
Key among these are forms of dependence created by the need to convince
someone who disagrees with you.  If you are advocating a position
in a debate with opposing sides leading to an eventual decision 
(for example, a Supreme Court case,
or a policy discussion on Wikipedia prior to a vote), then
your audience can be roughly divided into two groups: people who would
naturally tend to vote in favor of your position, 
and people who would naturally tend to vote against your position.
Principles of exchange theory indicate that in such situations,
you are more dependent on the people who would naturally vote against you,
and less dependent on the people who would naturally
vote
 for you,
since in order to accomplish your goal, you need to effect a
more substantial behavior change in the former group
\cite{emerson-power-dependence,kotter-dependence-mgmt,wolfe-power-negotiation}.
An important further point here is that
in our settings, 
participants can readily anticipate, either through dialogue
or advance knowledge, who is ``on their side'' and who is ``on the
other side,'' and so it makes sense to suppose that they are
aware of these dependence relations during the interaction.

Motivated by this, in the Supreme Court data
we will compare levels of coordination of lawyers toward 
\markhigh{unfavorable Justices} who 
(eventually)
vote against their side 
and toward \marklow{favorable Justices} who 
(eventually)
vote for their side;
there is more dependence and hence more of a power difference
in the former case.
In the Wikipedia data, 
we will compare levels of coordination of editors with others
who \markhigh{vote the opposite way} and with others
who \marklow{vote the same way}; here too, there is more
dependence and hence more of a power difference in the former case.
We should also note the exchange-theoretic principle that a dependence
relation affects both sides: $A$'s dependence on $B$ is expected
not just to affect $A$'s behavior in their interaction, but $B$'s as well.


\section{Linguistic style coordination}

As discussed earlier, we use {\em linguistic style coordination} 
to quantify the degree to which 
one individual \textit{immediately} echoes the linguistic style of the person they are responding to.
Here, linguistic style is quantified by a person's usage of certain
linguistic style \markers
(= categories of function words).
We first describe these \markers, then give formal definitions of coordination.

\subsection{Linguistic style \markers}
\label{sec:coord:marker}
We measure the linguistic style of a person by their usage of
categories of 
function words 
that have little
semantic meaning,
thereby marking style rather than content.

For consistency with prior work, 
we employed 
eight of the
nine LIWC-derived
categories \citep{liwc} deemed  to be
processed by humans in a
generally non-conscious fashion \cite{Ireland01122010}.
Our 
eight
{\em \marker}s are thus:
articles, 
auxiliary verbs, 
conjunctions, 
high-frequency adverbs, 
impersonal pronouns, 
personal pronouns,
prepositions, and
quantifiers 
(451 lexemes total).\footnote{
We discarded negation because it is sparse and seems to carry
semantic meaning.  \cite{Ireland01122010} also discarded some negations.}

\subsection{Coordination measures}
\label{sec:coord:formal}

\newcommand{\tone}{\ensuremath{u_1}\xspace}
\newcommand{\ttwo}{\ensuremath{u_2}\xspace}
\newcommand{\trand}{\ensuremath{u}\xspace}
\newcommand{\uone}{\ensuremath{a}\xspace}
\newcommand{\utwo}{\ensuremath{b}\xspace}
\newcommand{\UONE}{\ensuremath{A}\xspace}
\newcommand{\UTWO}{\ensuremath{B}\xspace}
\newcommand{\urand}{\ensuremath{b}\xspace}
\newcommand{\vrand}{\ensuremath{a}\xspace}

\newcommand{\EX}{\ensuremath{S_{\uone,\utwo}}\xspace}
\newcommand{\EXg}{\ensuremath{S}\xspace}
\newcommand{\EXgtarget}{\ensuremath{S_{\UONE,\urand}}\xspace}
\newcommand{\EXgsource}{\ensuremath{S_{\urand,\UTWO}}\xspace}

\newcommand{\markermath}{\ensuremath{m}\xspace}
\newcommand{\ind}[1]{\ensuremath{#1^\markermath}\xspace}
\newcommand{\indone}{\ind{{\cal E}_{\tone}}} 
\newcommand{\indtwo}{\ind{{\cal E}_{\ttwo\hookrightarrow\tone}}}
\newcommand{\Cm}{\ensuremath{C^{\markermath}}}
\newcommand{\Cq}{\ensuremath{C^{quant}}\xspace}
\newcommand{\avg}[2]{\langle #2 \rangle_{#1}}

Here we present a variation
and further analysis 
 of the measure 
introduced
in 
\cite{Danescu-Niculescu-Mizil+al:11a}, adapted 
to the setting of group conversations.

\vpara{Coordination with respect to a \marker} We start by defining the coordination of one person
\utwo towards another person \uone with respect to a specific linguistic style \marker \markermath. 
We want to quantify how much the use of \marker class \markermath
 in an utterance of \uone's \textit{triggers} the occurrence of \markermath in \utwo's {\em immediate} (meaning next)  reply to that utterance. 
 To put it another way, we want to measure how much \uone's use of \markermath in an utterance \tone increases the probability that \utwo will use \markermath in 
his reply \ttwo, where the increase is relative to \utwo's normal usage of \markermath in conversations with \uone.
We stress that we are thus looking at a more subtle phenomenon than
whether $\utwo$ uses articles (say) more overall when talking to
$\uone$: we want to see whether $\utwo$ is so influenced by $\uone$ as
to change their function-word usage in their very next reply.

Recall from \S \ref{sec:hyp} that we call \utwo the \textit{speaker} and \uone the \textit{target} of a conversational exchange
 $(\uone:\tone,\ \utwo:\ttwo)$, 
since $\uone$ is the target of $\utwo$'s reply when $\utwo$
speaks.
We say an utterance {\em exhibits} \markermath if it contains a word from
category \markermath. Let \indone be the 
event that utterance \tone 
(spoken to $\utwo$)
exhibits \markermath; similarly, let \indtwo be the event that
 reply \ttwo to \tone exhibits  \markermath.

Given a set
\EX of exchanges $(\uone:\tone,\ \utwo:\ttwo)$, we define the coordination of \utwo \textit{towards} \uone as:
\begin{equation}
\Cm(\utwo,\uone) = {P}(\indtwo ~|~ \indone) - {P}(\indtwo),
\label{eq:prob}
\end{equation}
where the probabilities are estimated over \EX, 
and where we require that at least one of $\uone$'s utterances exhibits \markermath in order for the first quantity to be defined.

\vpara{Properties} Eqn.~\eqref{eq:prob}  has several
interesting properties.  One non-obvious but important and useful
characteristic is that it is a function not only of $\utwo$'s
behavior, but also of $\uone$'s, because it can be shown
that \eqref{eq:prob}  lies in the interval
 $\left[ - \left(1 -P(\indone)\right), 1 -P(\indone)\right]$.

To see why  $\uone$'s behavior needs to be taken into account,  consider one
 extreme case: where every utterance of $\uone$  to $\utwo$ exhibits \markermath.  Then $\Cm(\utwo,\uone) = 0$ no matter
 what $\utwo$ does in response, which makes sense because we have no
 evidence that any (or no) usage of articles by $\utwo$ is done in
 response to what $\uone$ does --- we don't have any test cases to see
 what $\utwo$ does when $\uone$ doesn't employ a \marker.

Another extreme case is also illustrative:  where $\uone$ uses
\markermath only a few times when speaking to $\utwo$, and $\utwo$
uses \markermath when and only when $\uone$ does. Then, 
$\Cm(\utwo,\uone)$ approaches 1 as $P(\indone)$ approaches zero.  Again,
this makes intuitive sense:
it is very unlikely that $\utwo$ matching 
$\uone$ exactly on the few times $\uone$ used \markermath is due
merely to chance.

Another property of measure~\eqref{eq:prob} is that it is not symmetric, which fits the purpose of this study well,
since the power relations we want to investigate are also
asymmetric. See \cite{Danescu-Niculescu-Mizil+Lee:11a} for further
discussion on the asymmetry.

\vpara{Coordination towards a group} In the context of group conversations, we can extend this definition to coordination 
of a particular
speaker $\urand$
towards a \textit{group of 
targets} \UONE by simply modifying the set of exchanges on which the probabilities in (\ref{eq:prob}) are
estimated.  Specifically, given a set 
\EXgtarget of exchanges $(\vrand:\tone, \urand:\ttwo)$ involving initial utterances \tone of various 
targets $\vrand\in\UONE$ and replies \ttwo of \urand, the coordination of \urand to the group \UONE is:
\begin{equation}
\Cm(\urand, \UONE) = {P}(\indtwo ~|~ \indone) - {P}(\indtwo),
\label{eq:perusermarker}
\end{equation}
but
where this time the probabilities are estimated over \EXgtarget.  

{

We then define the coordination of one group of people \UTWO towards another group \UONE as
the average coordination of speakers in \UTWO to targets in \UONE:
\begin{equation}
\Cm(\UTWO, \UONE) = \avg{\urand \in \UTWO}{\Cm(\urand, \UONE)}
\label{eq:groupavg}
\end{equation}
By taking the macro (unweighted) average, 
our measure will not be dominated by a few active speakers in a dataset.

\vpara{Aggregated measures} 
It is important to note that in general, coordination is multimodal: it does not necessarily
occur simultaneously
for all \markers \citep{Ferrara:1991}, and
speakers may coordinate on some features but 
diverge 
on others \citep{Thakerar+al:1982}.   
Hence, we also use aggregated measures of coordination of \UTWO to \UONE  to
provide an overall picture of the level of coordination between 
the groups.

Ideally we want to simply compute $C(\urand, \UONE)$ as the macro-average of 
$\Cm(\urand, \UONE)$ across different \markers \markermath,
and then compute $C(\UTWO, \UONE)$ 
the
same way as in (\ref{eq:groupavg}). 
Recall, however, that $\Cm(\urand, \UONE)$ can only be computed if $\EXgtarget$
contains enough exchanges exhibiting \markermath to reliably estimate both probabilities in 
(\ref{eq:perusermarker}), which is not always the case for all people
with respect to all \markers.
For instance, some persons rarely use quantifiers,
leaving \Cq undefined in those instances.

We accounted for such ``missing values'' in three different ways,
resulting in three aggregated measures:
\begin{description}
\item[Aggregated 1]  Compute the ``ideal'' macro-average $C(\urand, \UONE)$  only for the persons \urand for whom $\Cm(\urand, \UONE)$ can be computed for all \markers;
ignore all the others.
This reduces the set of persons considered by the
aggregated measure, but provides the most direct measure (in the sense
that it does not rely on any particular ``smoothing'' assumptions as
the 
next two aggregated measures do).
\item[Aggregated 2]  For each person \urand, if $\Cm(\urand, \UONE)$ is undefined, we ``smooth'' it by using the group average $\Cm(\UTWO, \UONE)$ instead; this
measure considers everybody for which we can compute coordination for at least one marker, but assumes 
that
people in a given group share similar coordination behavior.
\item[Aggregated 3]  For each person \urand, we take the average only over the \markers for which $\Cm(\urand, \UONE)$ is defined; this is
equivalent to assuming that \urand would have exhibited the same level of coordination for the missing \markers as they did with other \markers. This aggregation also considers everybody for which we can compute coordination for at least one marker.  
\end{description}

\newcommand{\highpower}{G^{high}}\newcommand{\lowpower}{G^{low}}
\subsection{Formalization of the power hypotheses} Now that we have 
introduced a more formal definition of coordination between two groups of people,
we
formalize
the hypotheses introduced in 
\S\ref{sec:hyp} in terms of this definition.
If people in a group \markhigh{$\highpower$} have more power than people in a group \marklow{$\lowpower$}, and $U$ is a set of
arbitrary
 people, the power hypotheses can be rewritten as:

\begin{description}
\item \toone: $C(U, \markhigh{\highpower}) > C(U, \marklow{\lowpower})$
\item \fromone: $C(\markhigh{\highpower}, U) < C(\marklow{\lowpower}, U)$
\end{description}


\section{Empirical Investigation}
\label{sec:mainres}

Using the concepts and formalism introduced in the previous sections, we can now investigate the relation between linguistic coordination and power differentials in concrete conversational settings.  Specifically, we test whether the principle \powerhyp and the hypotheses \toone and \fromone 
introduced in \Section \ref{sec:hyp} can be  empirically confirmed in the two datasets described in \Section \ref{sec:data}.  
We begin by discussing power differences arising from status in 
Wikipedia (where our primary status distinction will be 
admins vs. non-admins) and in the Supreme Court (where our primary
status distinction will be Justices vs. lawyers).
After this, we consider power differences arising from dependence.

\subsection{Power from status: Wikipedia}
First, communication behavior on Wikipedia provides 
evidence for hypothesis \toone: users coordinate more 
toward
the (\higher) \admins than toward the \nonadmins (Figure
\ref{fig:status:toadmins}).\footnote{
The major explanatory factor for these results does not appear to be wholesale repetition
of phrases, even short ones.  We note,  for example,  that with
respect to the data used for computing conjunction coordination,
only 
0.7\% of the exchanges contain trigram repeats involving conjunctions and only 3.5\%
contain bigram repeats involving conjunctions; and  the difference in coordination levels
remains significant when exchanges with such repeats are
discarded. 
}

In the other direction, however, 
when comparing
\markhigh{admins} and \nonadmins as speakers, 
the data provides evidence that is initially at odds with \fromone: 
as illustrated in Figure \ref{fig:status:fromadmins}, \admins coordinate 
to other people {\em more} than \nonadmins do (while the hypothesis predicted that they would coordinate {\em less}).\footnote{
Note that the observations shown in Figure \ref{fig:status:toadmins} do not imply those in Figure \ref{fig:status:fromadmins}, nor vice-versa.  For example, the trend in 
Figure \ref{fig:status:toadmins}
does not change if we restrict the speakers to be only non-admins (or only admins).}
We now explore some of the subtleties underlying this result,
showing how it arises as a superposition of two effects.

\begin{figure}[t!]
\centering
\subfigure[Supporting \toone]{
  \includegraphics[height=2.0in,type=pdf,ext=.pdf,read=.pdf]{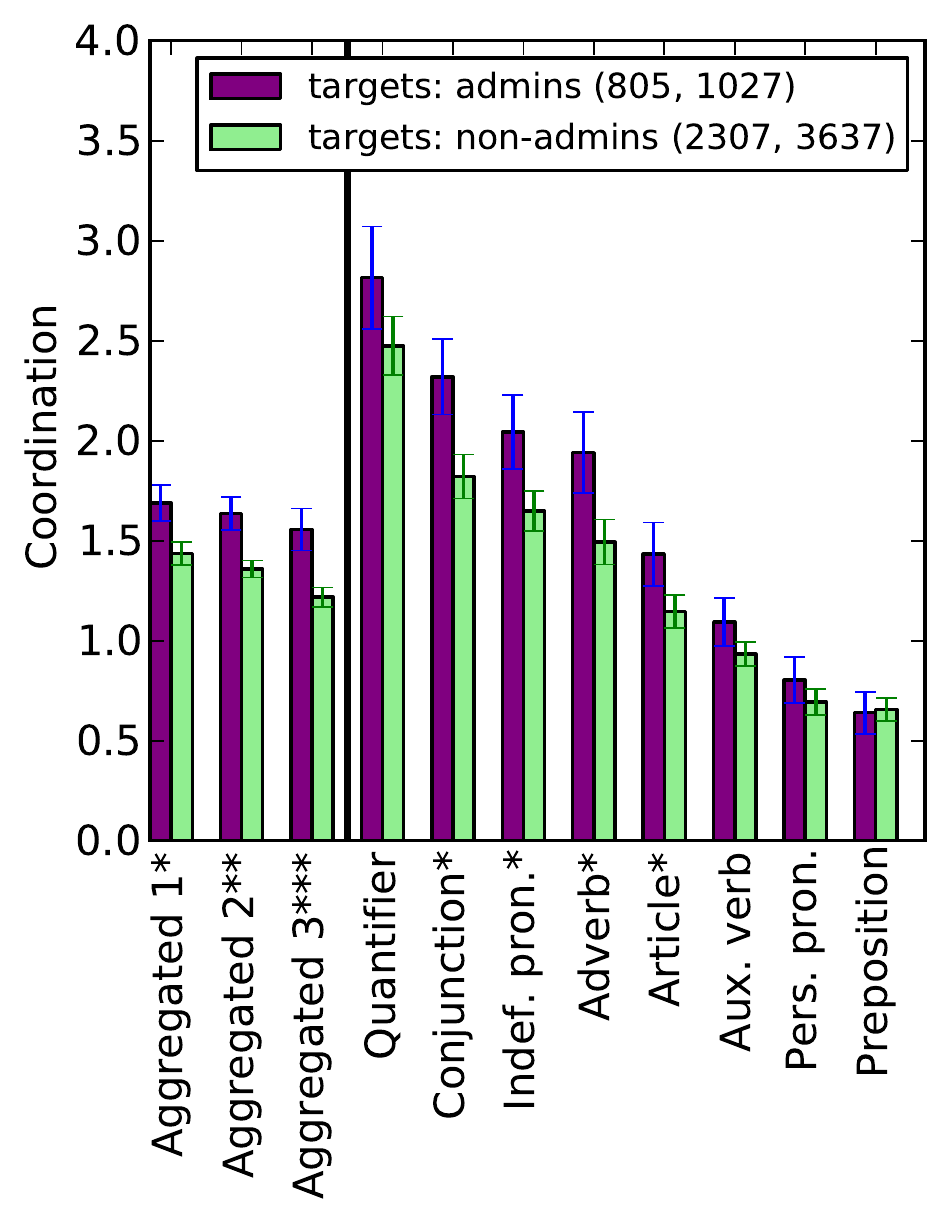}
  \label{fig:status:toadmins}
}
\subfigure[Contradicting \fromone]{
  \includegraphics[height=2.0in,type=pdf,ext=.pdf,read=.pdf]{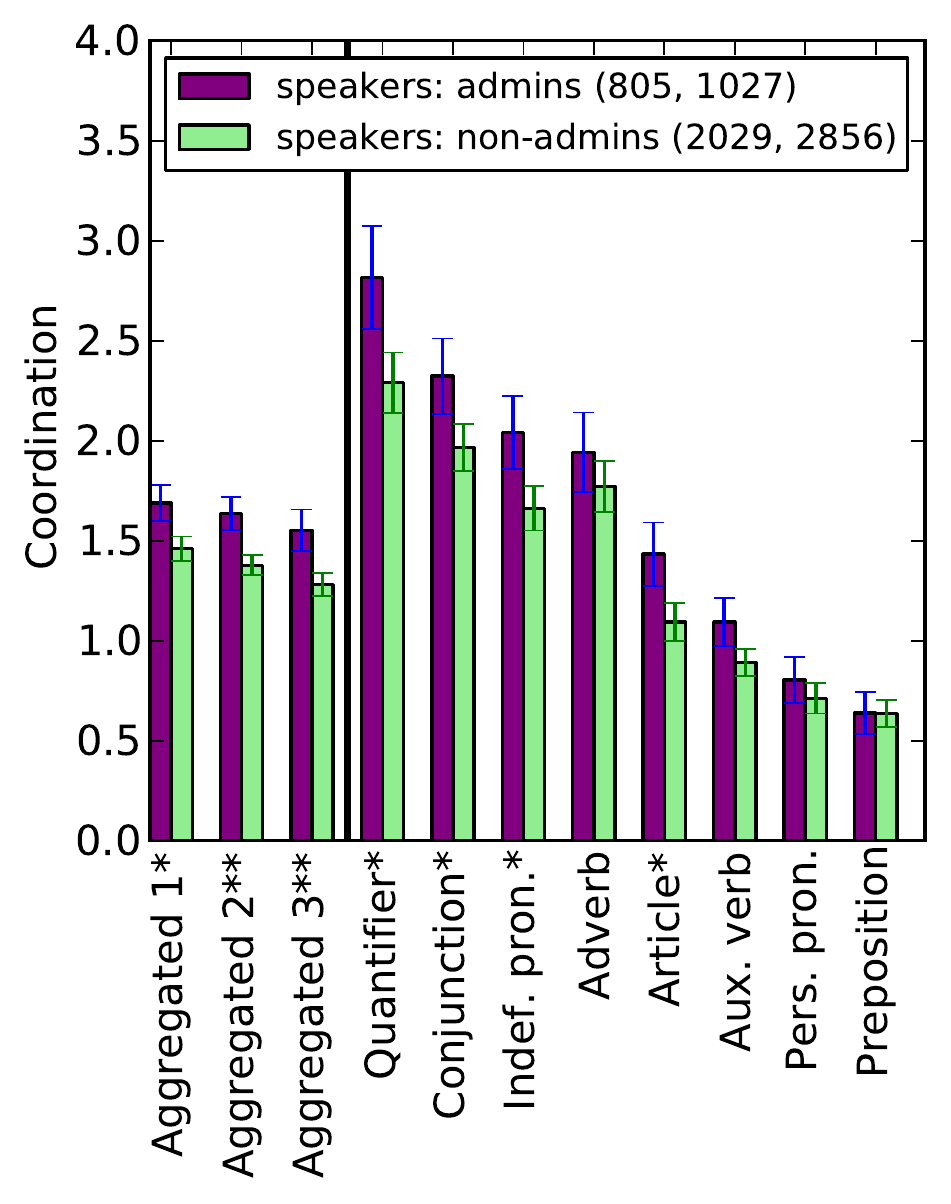}
  \label{fig:status:fromadmins}
}
\caption{Status and linguistic coordination: (a) Users coordinate
more towards \admins (\high) than towards \nonadmins (\low),
supporting  hypothesis \toone (indeed, significantly so in aggregate:
see later part of this caption). (b) On the other hand, \admins (\high) coordinate more than \nonadmins (\low) when replying to other people, 
contradicting hypothesis \fromone. \newline
\textcolor{red}{Note on all figures}: 
$*$'s on the x-axis
(e.g., ``Article*''
in (a)) indicate statistical significance,
independent
  t-test: *=``p<0.05'',**=``p<0.01'',***=``p<0.001''. 
Next to each legend label, in parentheses, are: the number of users
for
Aggregated 1 
(i.e., the users for which we can compute coordination for all
\markers) and the total number of users 
for
Aggregated 2 and
3 (i.e., the users for which we can compute coordination for at least
one \marker).
``Error bars'' do  {\em not} indicate standard error (we already
marked statistical significance with stars) but rather  give an idea
of 
how coordination values vary via
the standard deviation,
estimated by bootstrap
resampling \cite{Koehn:ProceedingsOfEmnlp:2004}.
The y-axis values are 
reported as percentages 
(i.e., multiplied by 100) 
for clarity.
}
\label{fig:status}
\end{figure}

\vpara{Personal characteristics: Hypothesis \fromtwo}
One possible explanation for the inconsistency of our observations
with \fromone is the effect of personal characteristics suggested
in Hypothesis \fromtwo from \S \ref{sec:hyp}.
Specifically, admin status was not conferred arbitrarily on a set of users;
rather, admins are those people who sought out this higher status and
succeeded in achieving it.  
It is thus natural to suppose that, as a group, they may have
distinguishing individual traits that are reflected in their level
of language coordination.

Fortunately we can extract rich enough data from Wikipedia that 
it becomes possible,
to a significant extent, to separate the effect of status from 
these individual traits, establishing that both effects play a role.
Our separation of these effects is based on the fact that
status can change abruptly, while personal characteristics, though mutable,
are more stable over time.
On Wikipedia, 
status changes are well documented, as they can occur only through
an election process instigated by 
requests for adminship (RfAs).  
When we compare the set of \adminstobe --- future admins before they were
promoted via their RfA --- with \nonadmins, 
Figure \ref{fig:adminsbefore} shows that the same differences in
language coordination were already present in these two populations ---
hence, they are not an effect of status alone, since they were
visible before the former population
experienced a status upgrade.

\begin{figure}[t!]
\centering
\subfigure[Before RfA elections (\fromtwo)]{\includegraphics[height=2.0in,type=pdf,ext=.pdf,read=.pdf]{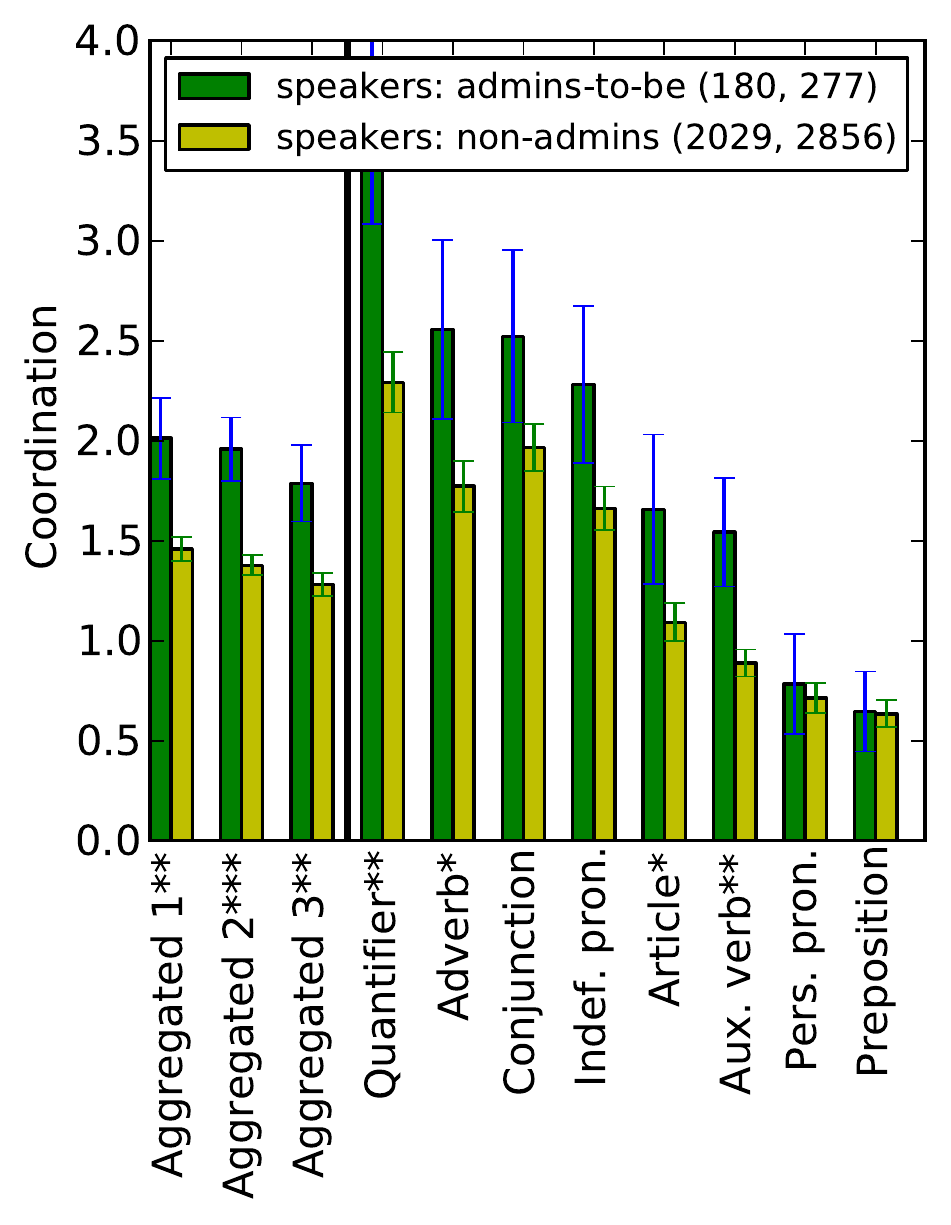}\label{fig:adminsbefore}}
\subfigure[In RfA discussions (\fromtwo)]{\includegraphics[height=2.0in,type=pdf,ext=.pdf,read=.pdf]{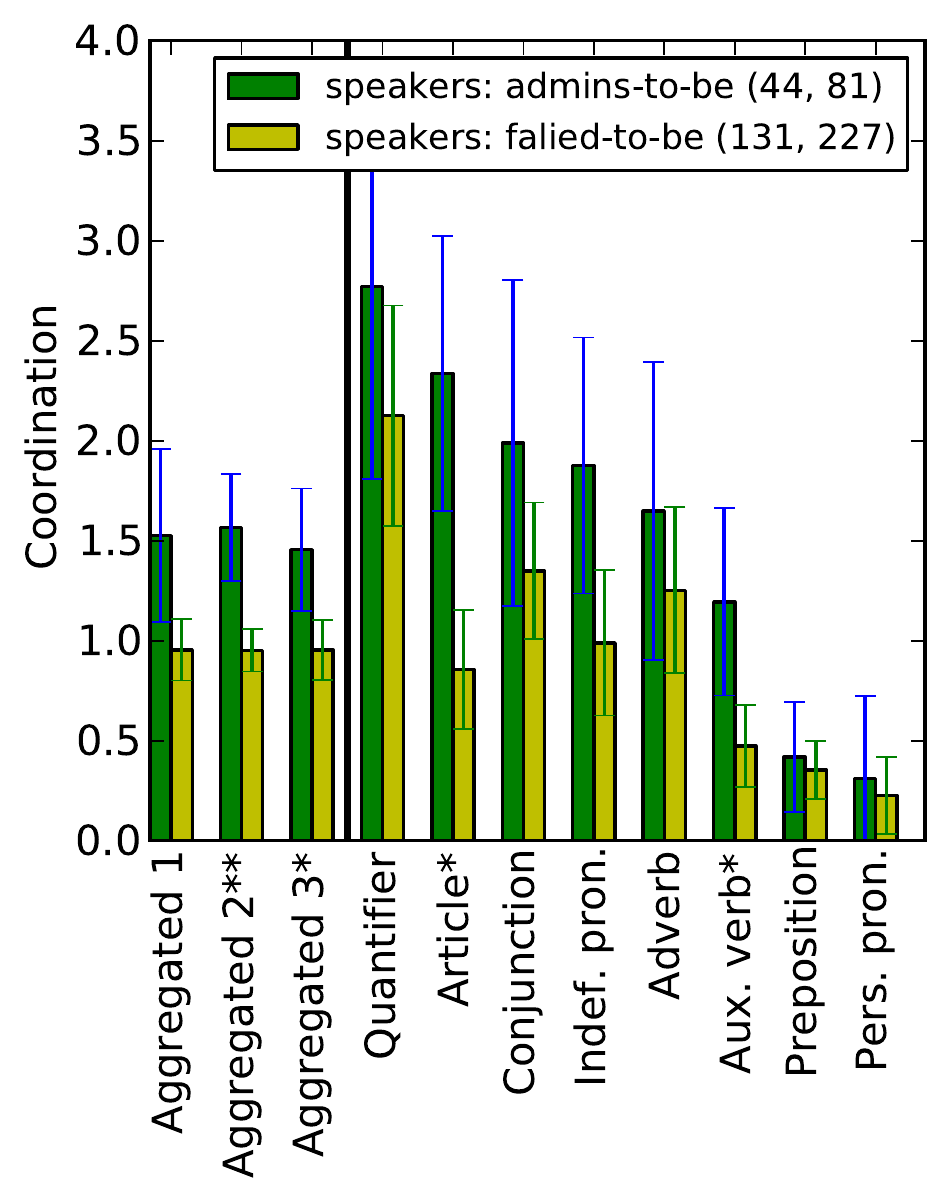}\label{fig:elections}}
\caption{Same-status comparisons (supporting a ``winner'' personality hypothesis): (a) \adminstobe coordinate more than those who remain \nonadmins thoughout; (b) 
during adminship elections (RfAs), \adminstobe coordinate more than \failedtobe. 
 }
\end{figure}
\vspace{1cm}

Can we separate the effects of ambition from success?  Yes, because we
can look at 
differences in coordination between users who were promoted 
(\adminstobe), and
those who went through the RfA process but were denied admin status
(\failedtobe).
Both \adminstobe and \failedtobe had the ambition to become
admins, but only members of the former group
succeeded.
We investigate coordination differnces between these two groups during 
a period when their adminship ambitions are arguably most salient:
during the discussions in each user's own RfA process.
Figure \ref{fig:elections} shows that even in the conversations they
had on their RfA pages, the \adminstobe were coordinating more to the
others than the \failedtobe, providing evidence for a strong form
of Hypothesis \fromtwo.

\vpara{Revisiting status: Hypothesis \fromoneprime}
We now return to the issue of status, and describe a method of
partially controlling for personal characteristics so as to evaluate
the following modification of Hypothesis \fromone:

\begin{description}
\item \fromoneprime. When controlling for personal characteristics, \high people coordinate less than \low people.
\end{description}

To study \fromoneprime, we create two populations for comparison:
the interactions of each \markhigh{admin} before his or her promotion via RfA
(i.e., when they were \adminstobe),
and the interactions of each \markhigh{admin} 
after his or her respective promotion.
Figure \ref{fig:beforeafter} shows how the resulting comparison
confirms \fromoneprime: \adminstobe
decrease their level of coordination once they gain
power.\footnote{Note that the trend shown in Figure
\ref{fig:beforeafter} is maintained when
considering the exact same users in both groups (i.e., excluding the
users which did not have enough conversations both before and after
adminship). Also note that we allow a time buffer 
of
a month after the
RfAs between the two sets of conversations we compare.} 
Interestingly, the reverse seems to be true
for \failedtobe: after failing in their RfAs --- an event that arguably
reinforces their failure to achieve high status in the community 
--- they coordinate more
(
p-value 
< 
0.05; 
figure omitted due 
to space limitations).

In addition, we can employ 
 status change  to reinforce \toone in a 
setting that controls for personal characteristics:
we find that users coordinate
more to \admins after promotion than 
when they were \adminstobe (p-value<0.05).

Finally, in Figure \ref{fig:statuschange}, we investigate 
how quickly the change in status is reflected
in the communication behavior of the users involved.
In addition to the monotonic changes in coordination levels over time,
and in the hypothesized directions, 
it is interesting to note that the most
dramatic change in coordination is visible in
the second month after the change in status occurred. This suggests
a period
of
acclimation
 to the newly gained status, both for the
person that undergoes the change and for those witnessing it.

\begin{figure}[tb!]
\centering 
\subfigure[Supporting \fromoneprime]{
 \includegraphics[height=1.95in,type=pdf,ext=.pdf,read=.pdf]
 {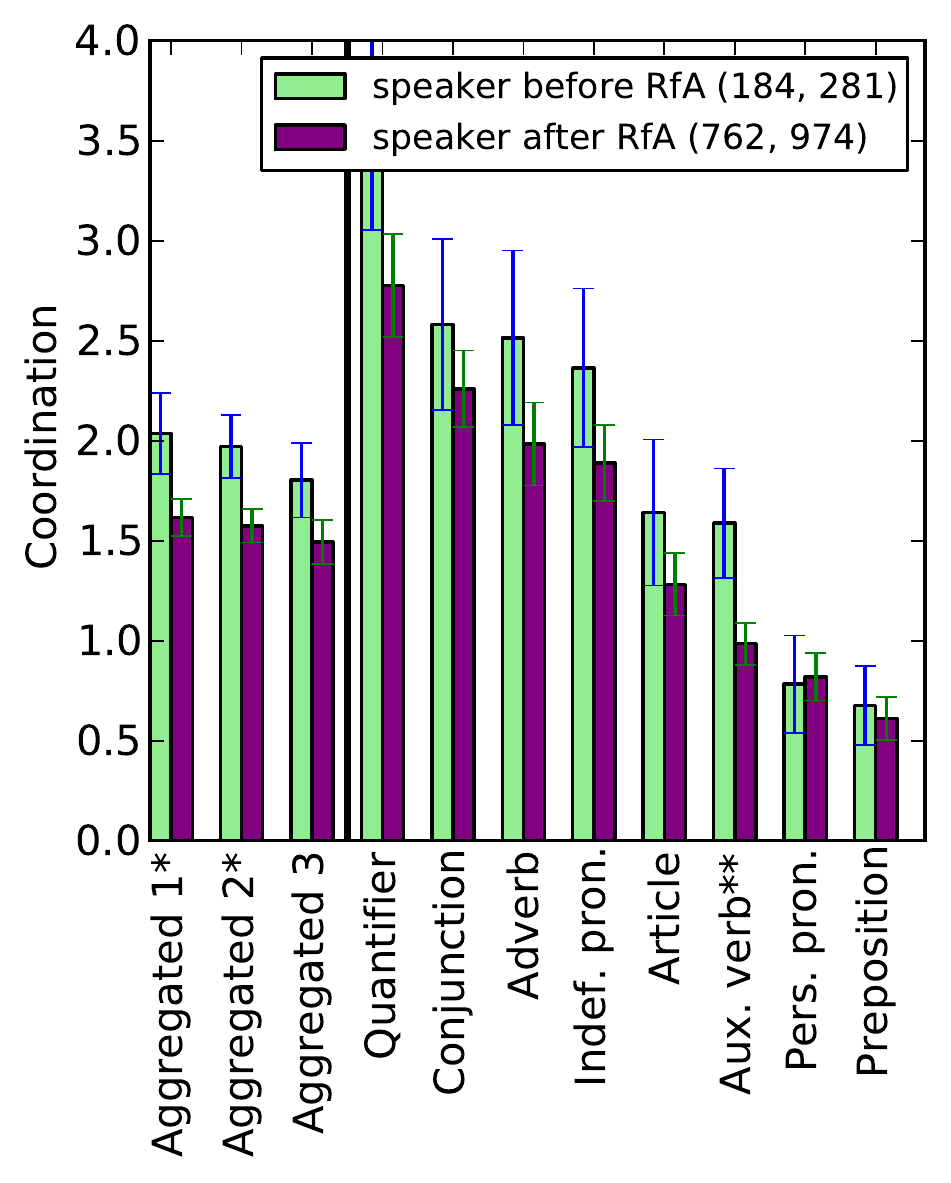}
 \label{fig:beforeafter}
}
\subfigure[Timed effect of  \textbf{status change} (\powerhyp)]
{
  \includegraphics[height=1.95in,type=pdf,ext=.pdf,read=.pdf]{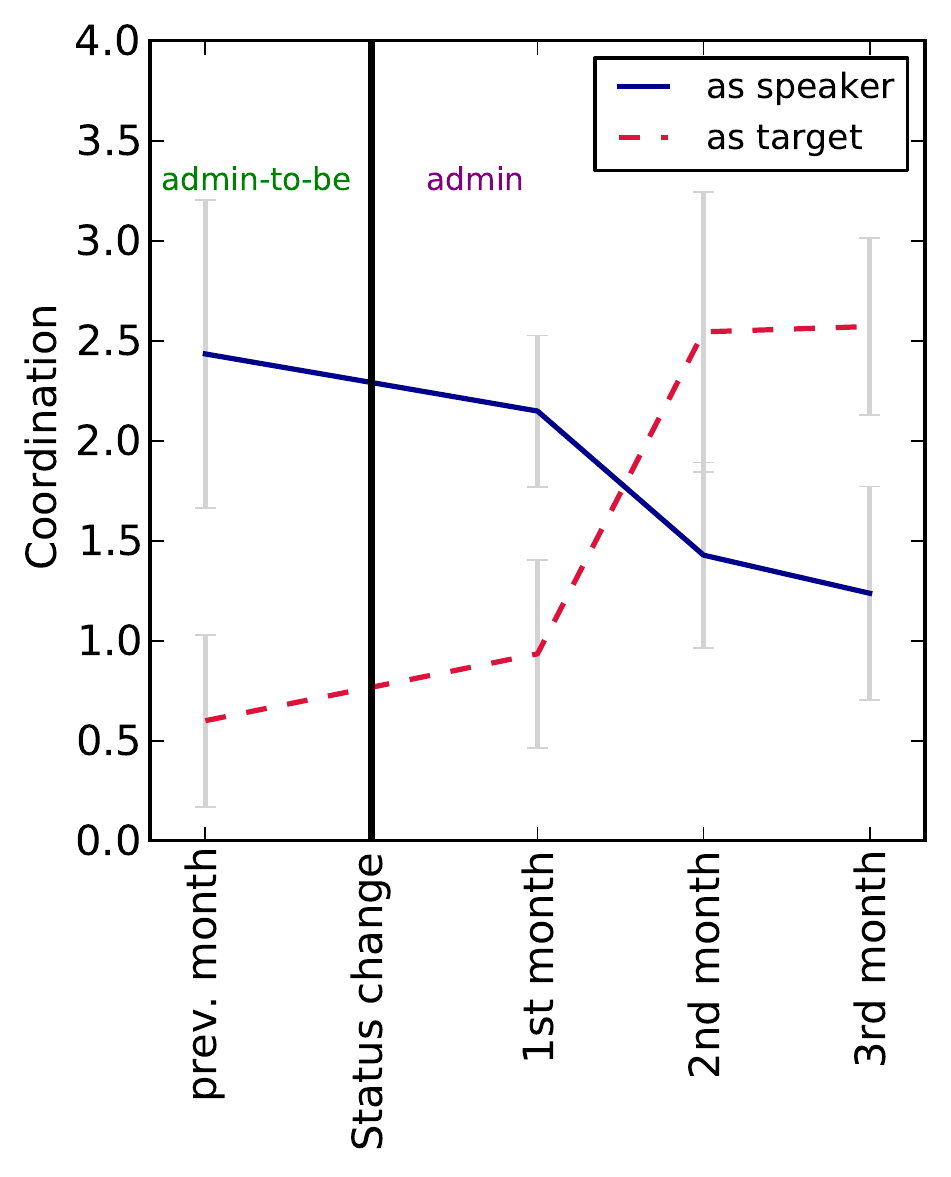}
  \label{fig:statuschange}
}
\caption{Effect of status change.  (a) \adminstobe coordinate less \markhigh{after they become admins}; 
(b) Aggregated 1 coordination of the user (as speaker) and,
respectively, towards the user (as target) 
before and after status change occurs through RfA.}
\end{figure}

\subsection{Power from status: Supreme Court}

In the setting of the \court, status differences are extremely salient
and do not suffer from the correlations that added complexity to the study of
\fromone in its original form.  Also, conversations during
the oral arguments (almost) always are between a \justice and a
\lawyer.  Thus, our basic finding can be expressed succinctly in
Figure \ref{fig:court:status}, which shows significantly 
more coordination from \lawyers to \justices than 
vice versa.\footnote{
Throughout, we consider each appearance of a given Justice or lawyer in a
different case as a separate entity, which allows for different
behaviors in different cases and increases the number of datapoints.
}.

In the \court setting we can also study finer-grained status distinctions,
to see if these too are manifested in language coordination
differences.
Indeed, 
in concordance with \toone, we observe that
lawyers coordinate significantly more toward the \markhigh{Chief
Justice} than toward the \marklow{Associate Justices} (p-value<0.01).%

\begin{figure}[t!]
\centering
{\includegraphics[height=2.2in,type=pdf,ext=.pdf,read=.pdf]{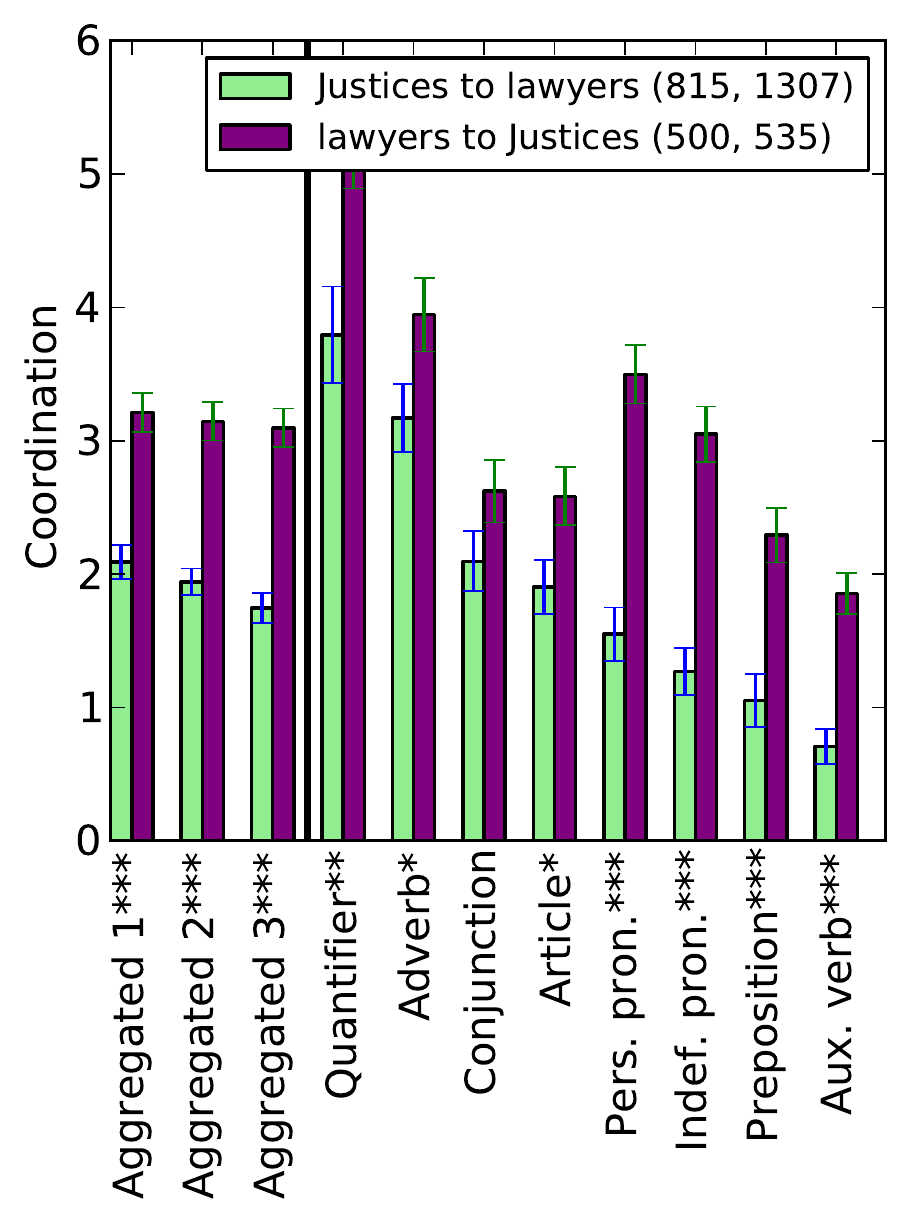}}
\caption{
Lawyers coordinate more \markhigh{to} \justices than conversely.}\label{fig:court:status}.
\vspace*{-0.3cm}
\end{figure}

\newpage

\subsection{Power from \dependence}
\begin{figure}[t!]
\centering
\subfigure[Dependence: \toone]
{\includegraphics[height=2.0in,type=pdf,ext=.pdf,read=.pdf]{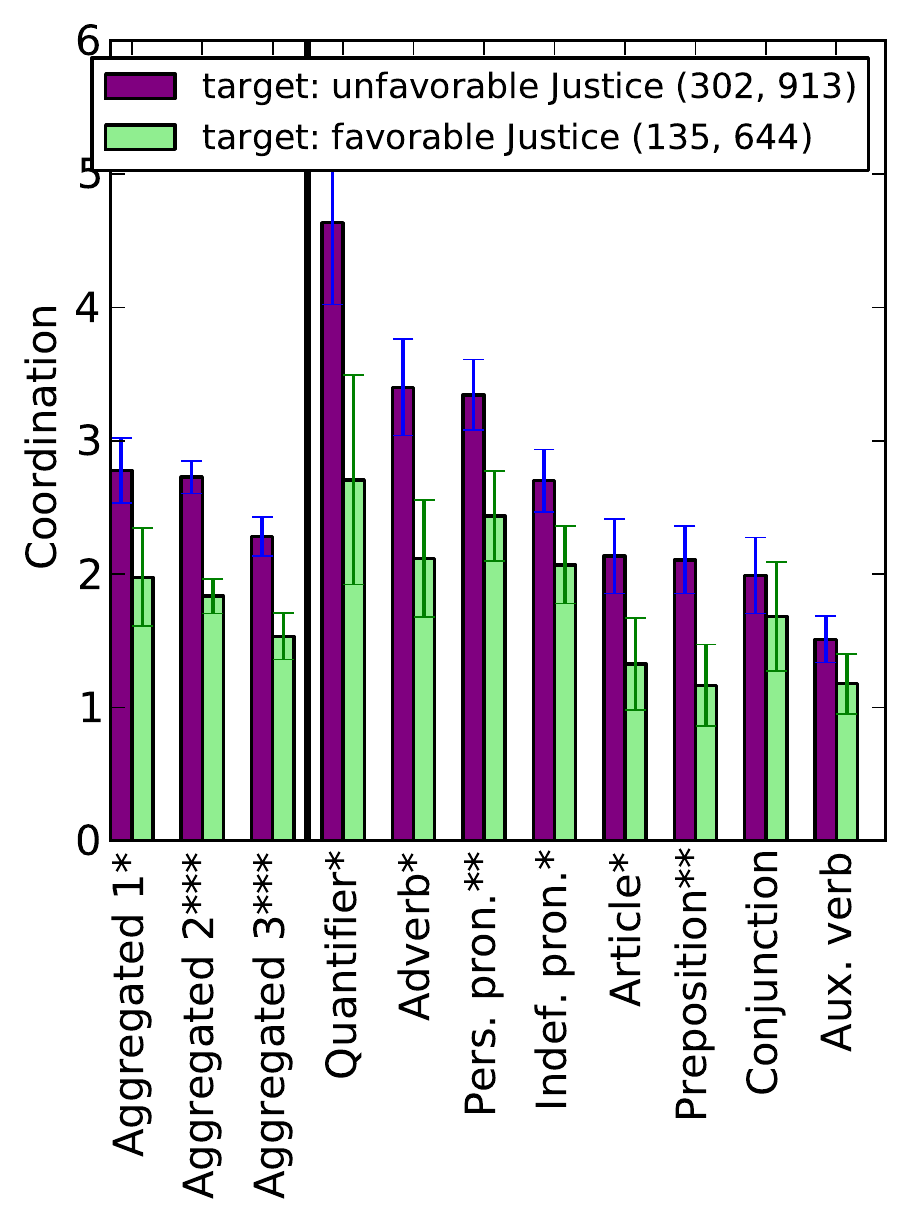}
\label{fig:tofavorable}}
\subfigure[Dependence: \fromone]
{\includegraphics[height=2.0in,type=pdf,ext=.pdf,read=.pdf]{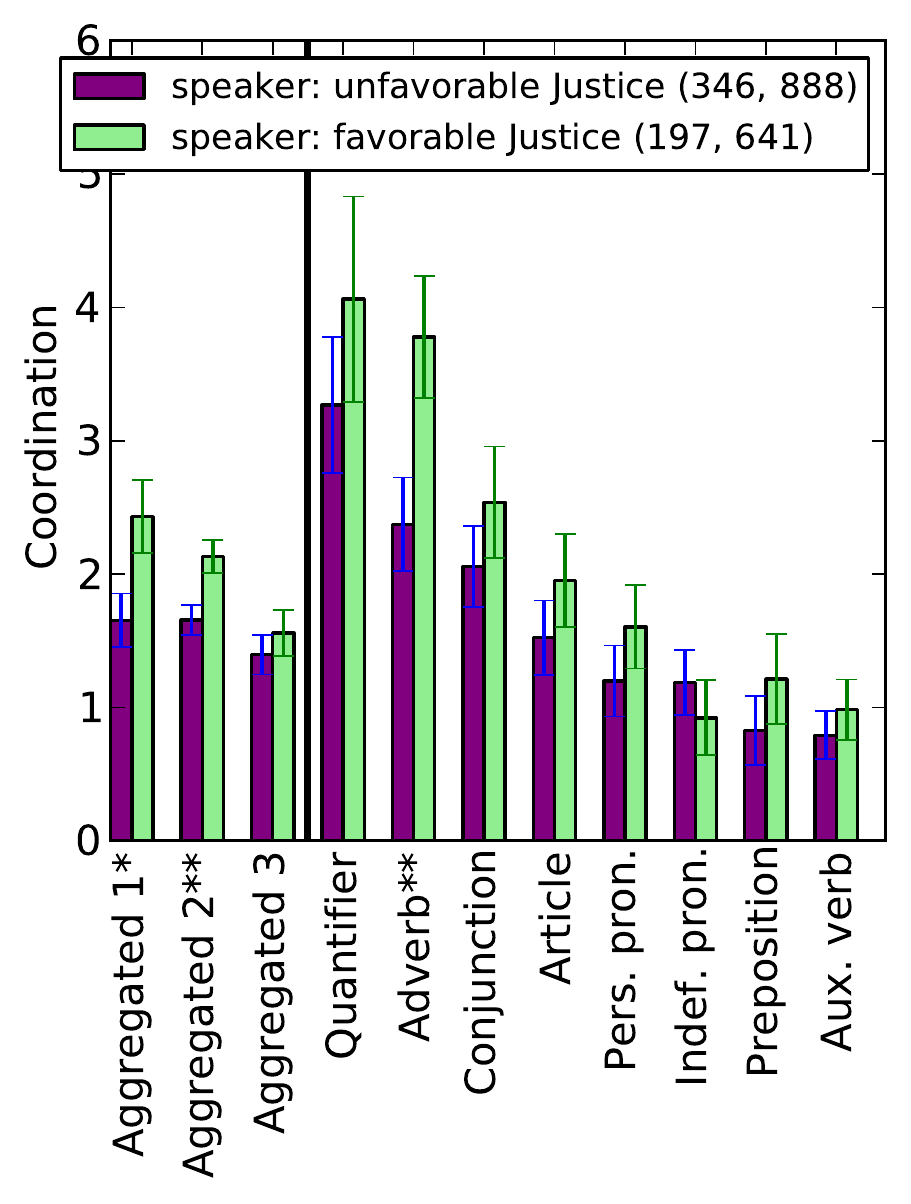}\label{fig:favorable}}
\caption{Dependence and linguistic coordination: (a) lawyers adjust their coordination level according to whether the Justice is \markhigh{unfavorable} or \marklow{favorable}, supporting \toone;  (b) \favorables coordinate more than \unfavorables (\fromone).}
\end{figure}

As noted in 
\S \ref{sec:data}, we can study power differences
based on dependence --- even for fixed levels of status difference ---
using the exchange-theoretic principle that the need to convince someone
who disagrees with you creates a form of dependence
\cite{emerson-power-dependence,kotter-dependence-mgmt,wolfe-power-negotiation}.
Moreover, this power difference is predicted to be 
felt by both sides ---
the side with lower power and the side with higher power.

In the case of
lawyer-Justice interactions, 
let us define the Justice to be \marklow{favorable} to the lawyer if he or she
ends up voting on the lawyer's side, and \markhigh{unfavorable} if he or she
ends up voting against the lawyer's side.
It is 
well understood that 
the Justices often come into the case with a general leaning
toward one side or the other based on their judicial philosophy
--- this has been acknowledged for example in interviews with
members of the Court \cite{scalia} --- and lawyers
through their preparation for the case will come in with knowledge
of these leanings. Hence it is reasonable to suppose that the 
\marklow{favorable}-\markhigh{unfavorable} distinction will be salient to the 
interaction during oral arguments.

And indeed, 
Figures \ref{fig:tofavorable} and \ref{fig:favorable}
show that the power differences
created by this form of dependence are reflected in the amount
of coordination, in both directions.
First, lawyers coordinate more toward 
\markhigh{unfavorable Justices} (on whom they
are more dependent) than toward \marklow{favorable Justices}, 
in keeping with \toone.
Second, \markhigh{unfavorable Justices} coordinate
less toward lawyers than \marklow{favorable Justices} do, 
in keeping with \fromone.
Given the formal framework of {\em impartiality} that characterizes the 
Justices's behavior at the \court,
it is intriguing to see the
undercurrent of language coordination differences nevertheless hinting at
their eventual decision.

We see a similar effect of dependence on coordination 
in the context of discussions with opposing sides on Wikipedia.
During RfAs, one voter may try to change the opinion of voters
on the other side who have already cast their vote.
(Changing your vote during the RfA process is allowed, and hence
there is an incentive to convince voters to consider this.)
Users coordinated more when engaging with \markhigh{users on the opposite
side} than with \marklow{those who voted the same way}
(p-value<0.05; for space reasons we omit the figure).
This finding too, via the arguments about opposing sides and dependence,
supports the general power-coordination principle
\powerhyp.


\section{Cross-Domain Analysis and Interaction among Hypotheses}

\subsection{Coordination as a Cross-Domain Feature}
\label{sec:exp:pred}

Part of the motivation for studying 
the relation between coordination and power is that the 
principles that govern this relation
appear to be domain-independent.
Here we perform a set of analyses to show that coordination
features do generalize across our two domains 
more effectively than other text-based features
for the problem of inferring power. 
We find that
indeed, compared to the other features we consider, they are
the only set of features to display any non-trivial generalization.

Our analysis is based on the following prediction task:
for a given pair of 
different-status people \preda and \predb
who have engaged in conversations with each other, we predict whether \preda has the higher status.
In this setting, a random guess baseline would achieve 50\% accuracy.
We stress, however, that this prediction task is primarily a means 
to assess cross-domain generalization, i.e., not as a free-standing task in itself.
Indeed, the best achievable performance on this status-prediction task appears 
to be quite domain-dependent. In some domains such as the \court,
idiosyncratic cues in text usage (e.g., lawyers begin their sentences with
stylized lead-ins, such as ``Your honor'', that clearly mark them as lawyers, not Justices)
enable almost perfect performance when these cues are available as features.
In other domains, such as Wikipedia, an informal evaluation using
two human annotators familiar with the domain produced only $70\%$ accuracy (and an inter-annotator agreement of only $80\%$).
Thus, our interest is not in whether coordination features achieve
the best within-domain performance, but in whether they are particularly
effective at generalizing (as we indeed find them to be).

\vpara{Experimental setup}
Let \reply{\preda} be \preda's replies to \predb, and \reply{\predb} be \predb's replies to \preda, and \length{S} 
be the average length of all utterances in the set $S$.  Let \sfmath be the set of 8 stylistic \markers introduced in \S \ref{sec:coord:marker}.
We define the following sets of features used as input to an SVM classifier:
\begin{itemize}

\item \coordfeature: binary features indicating, for each $m \in \sfmath$ as well as for Aggregated 1\footnote{We only considered pairs of participants for which enough data was available to compute coordination on all stylistic features.}, whether \preda coordinates more to \predb than \predb to \preda on $m$
\item \baselinefeature: 
frequency of each \marker $m \in \sfmath$  in \replya and, respectively, in \replyb; also, \length{\replya}, \length{\replyb}. We use this feature set to examine whether style alone is predictive on its own, or whether specifically stylistic {\em coordination}  is key
\item \bowfeature: frequency of each word in \replya, frequency of each word in \replyb, $L_2$-normalized 
\end{itemize}

For experiments on the Wikipedia data, which we denote as \wikidatashort, we considered 
\textit{(admin, non-admin)} pairs 
(for conversations occurring after the admins were elected).
For the \court dataset (\courtdatarestricted), we considered \textit{(Justice, lawyer)} pairs\footnote{In order to focus on the conversational exchanges and avoid exchanges in which the lawyers formally introduce their case, we considered only cases where the length difference between the two utterances
were fewer than 20 words.}.

For {\em \indomain} experiments, we report average accuracy over cross-validation 
within the same domain (i.e., training and test corpora are both \wikidatashort or
\courtdatarestricted); for {\em \crossdomain} experiments, we train on one domain and test on the other.

\vpara{Results}
Table \ref{tab:pred} summarizes the results. 
We find 
that \coordfeature are the only ones to perform
statistically significantly better than random guessing 
in the \textit{\crossdomain} settings ---
the other classifiers simply learn cues that are idiosyncratic to
their training data, and fail to generalize.  
(Note for example 
that the bag-of-words method picks up on the near-perfect
lexical
cues marking lawyers in the \court data, but this method performs worse
than random guessing when applied to  the other domain.)

Even looking at the \textit{\indomain} tasks --- which were not
our primary focus here --- we find that 
\coordfeature are the only ones that perform statistically
significantly better than random guessing on \textit{both} datasets.

\begin{table}
\begin{tabular}{r|cc|cc}
&\multicolumn{2}{c|}{\indomain } &\multicolumn{2}{c}{\crossdomain} \\
Training corpus &\wikidatashort &\courtdatarestricted &\courtdatarestricted  &\wikidatashort\\
Test corpus &\wikidatashort &\courtdatarestricted &\wikidatashort &\courtdatarestricted\\ \hline
\coordfeature (9 altogether)& \textbf{57.7}	&\textbf{70.4}		&\textbf{57.1} &\textbf{55.0} \\	
\baselinefeature (18 altogether)& \textbf{59.2}	&51.4	&50.0 &51.9 	 \\
\bowfeature (20,000 altogether)& 51.4 &\textbf{99.5} &  45.2 & 40.1\\
\end{tabular}
\caption{
Prediction accuracy for SVM's using various feature sets. Cross-domain
results are
in the 
right-hand 
two
columns. Bold = 
results significantly
better than chance.}
\label{tab:pred}
\end{table}


\subsection{Interactions among Hypotheses \powerhyp and \fromtwo}

In \S \ref{sec:mainres} we saw that the interaction between
personal characteristics (which form the basis for
Hypothesis \fromtwo) and power differentials (which form
the basis for Hypothesis \powerhyp) can lead to complex effects.
Here we consider two cases where this interaction raises
interesting issues, and point to open questions in the 
analysis of coordination.

An individual's level of social engagement is one type of
personal characteristic that interacts with coordination and power.
As a simple proxy for social engagement, for purposes of discussion
here, we consider the volume 
of communication the individual engages in.
As we noted in \S \ref{sec:intro}, simple volume measures such as this 
do not seem to readily yield domain-independent information about power,
since they vary considerably across domains --- in some domains the
powerful people talk a lot, and in others they talk relatively little.
For example, when people are promoted to \markhigh{admin} status, their
volume of communication goes up while (as we have seen) their
level of coordination goes down.
On the other hand, \marklow{lawyers} talk more than \justices
in the \court data, and (again as we have seen) they also coordinate more
in the \lawyer-\justice interactions.

However, if we restrict attention to a fixed sub-population within
a given domain, there are interesting connections between coordination
and volume that suggest further questions.
In particular, on Wikipedia we consider 
the number of replies posted by a user on talk-pages as a measure
of communication volume, and hence a proxy for their level of
social engagement on the site.
We compared users in the top 1/3 of the sorted order by communication volume
with users in the bottom 1/3, finding that
users with higher numbers of replies are more likely to coordinate to others (p-value<0.05). %
We observed the same effect when we compared the communication
volumes of users with the same status:
among admins, users with more communication are also more likely to coordinate, %
and the same trend holds among non-admins.
Similar effects also hold for other measures of communication volume.
Again, we note that other domains (such as the \court) show an inverse relation
between volume and coordination in the communication transcripts,
and so it is an interesting question to identify the principles
that determine how this relationship plays out in different settings.

We also consider a second basic example that
raises an interesting challenge
for distinguishing between Hypotheses \powerhyp and \fromtwo:
the effect of gender on coordination, using the
fact that gender information is available for participants 
in the \court dataset.

\begin{figure}[ht!]
\begin{tabular}{ll}
  \includegraphics[height=1.95in,type=pdf,ext=.pdf,read=.pdf]{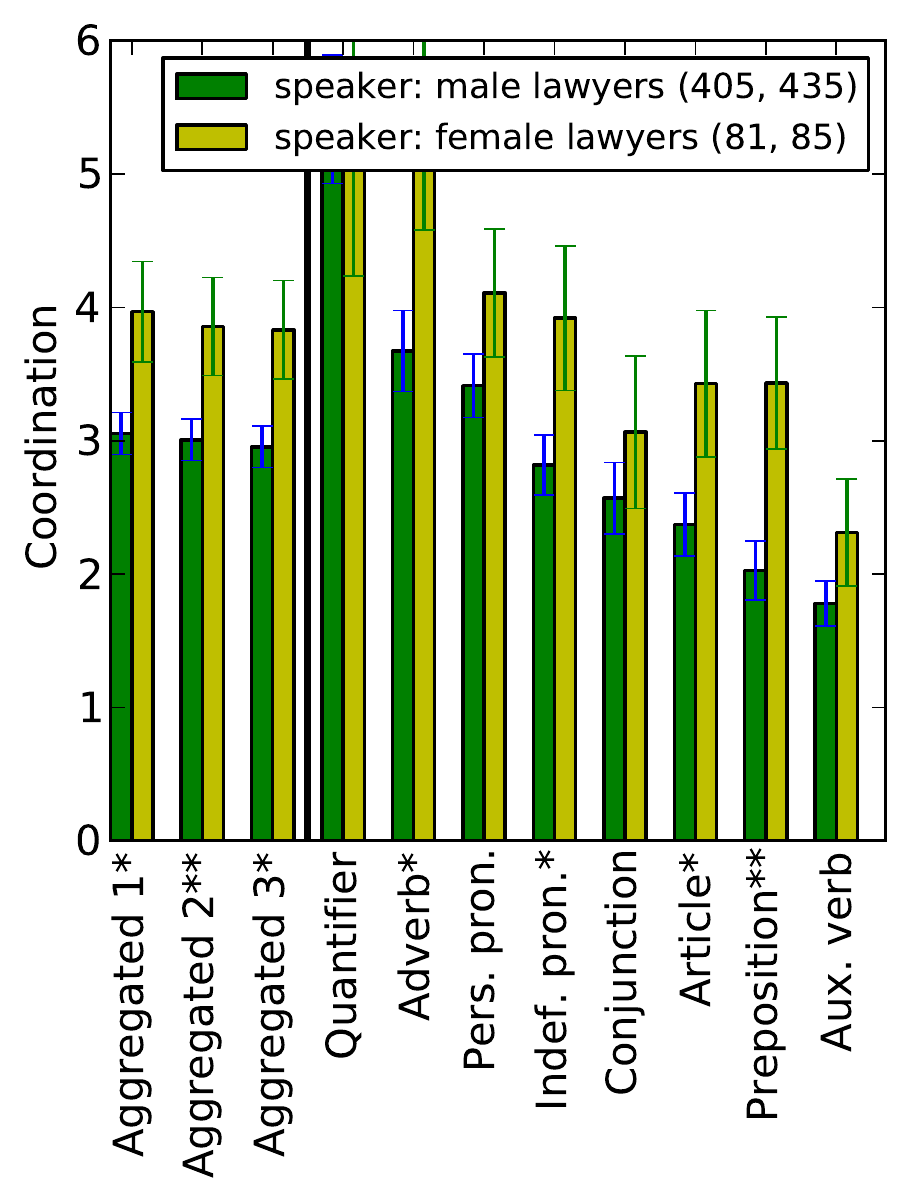} &
 \includegraphics[height=1.95in,type=pdf,ext=.pdf,read=.pdf]{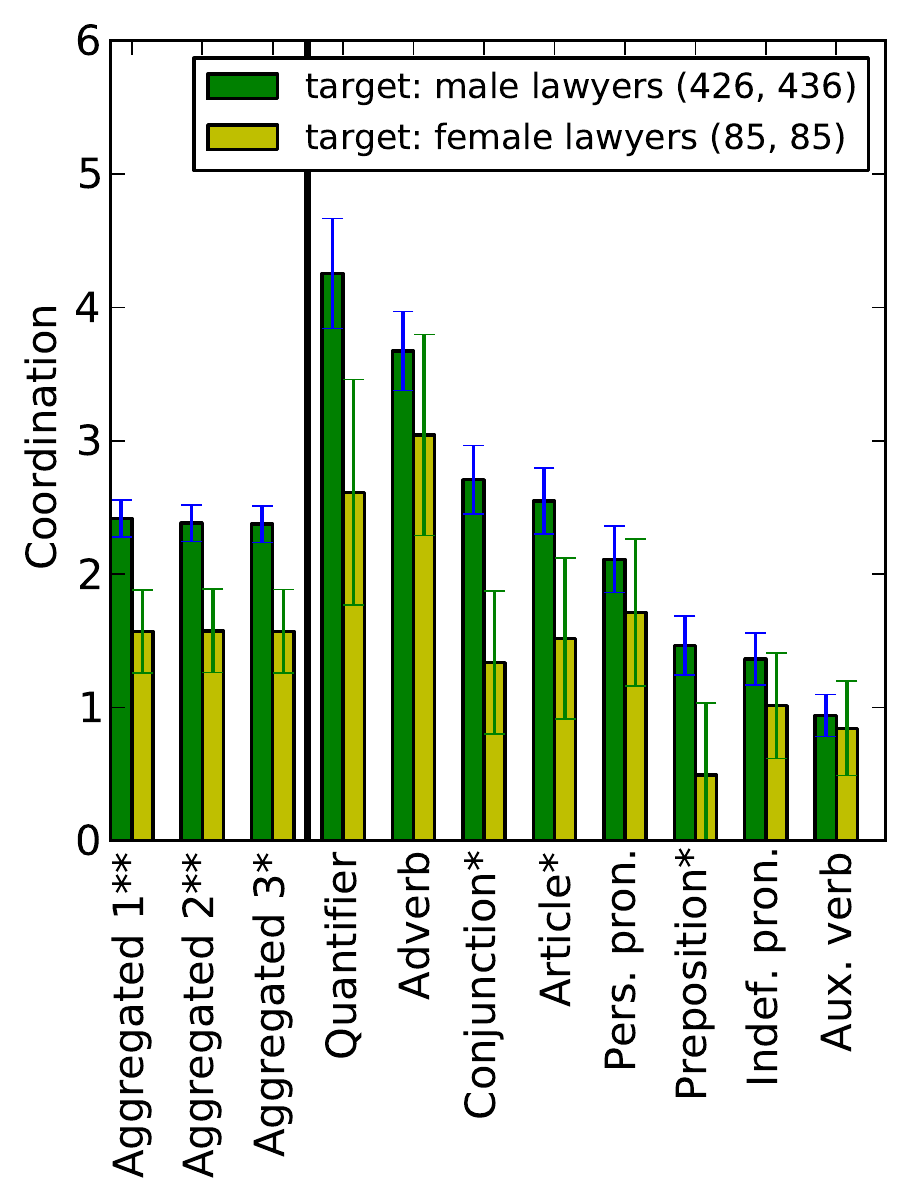}
  \\
{\small (a) gender differences in speakers} & {\small (b) gender differences in targets} \\
\end{tabular}
\caption{\label{fig:gender} Gender differences}
\end{figure}

The main finding here, in Figure \ref{fig:gender}, is that 
overall female lawyers coordinate more than male lawyers when talking 
to Justices,
and correspondingly, 
Justices coordinate more towards male lawyers than towards female lawyers.
Given the extensive history of work exploring 
culturally-embedded status and power differences based on gender
\cite{berger-sct-book,thye-status-value},
one interpretation of this finding is directly 
in terms of Hypothesis \powerhyp.
However, since it is also potentially related to 
theories of gender-based communication differences
\cite{herring-gender-power}
and even gender-based language adaptation differences \cite{Otterbacher:ProceedingsOfCscw:2012},
the question of separating Hypotheses $\powerhyp$ and $\fromtwo$
becomes 
challenging here.
We think it is a promising possibility
that language coordination effects may be able to 
serve as a lens through which to measure many similar kinds of 
distinctions in both on-line and off-line conversational settings.


\newpage
\section{Further Related Work}
\label{sec:relwork}

In the opening sections, we have discussed some of the ways in which
earlier work used text content to analyze on-line networks
\cite{hub2004,livne2011party,Choudhury2010,Menczer,Gilbert2009},
as well as background on language coordination and 
the exchange-theoretic notions of power from status and dependence.
Here we discuss some further work that is related to the general
issues we consider here.

\vpara{Power and structural features}  
There has been extensive work on using structural features, rather
than language, to infer notions of ``importance'' in networks,
both in the literature on social networks \cite{wasserman-faust}
and on the Web \cite{chakrabarti-mining-web}.  Recent work has
also studied the inference of status from on-line social network features
\cite{guha-trust,leskovec-www10}.

\vpara{Power and language}
The relation between linguistic coordination\footnote{For
  brevity, we exclude
studies
 of the effects of 
  status on other types of coordination, such as pitch and vocal features, which are absent from textually-manifested
  discussions (see 
  \cite{Giles:EngagingTheoriesInInterpersonalCommunicationMultiplePerspectives:2008}
  for a survey) or on related phenomena such as 
information-density matching \citep{Aronsson:JournalOfLanguageAndSocialPsychology:1987}.}}
and status has 
mostly been examined in small-scale contexts: 15 Watergate transcripts  \cite{Niederhoffer+Pennebaker:2002a}, 40 courtroom cases \cite{Aronsson:JournalOfLanguageAndSocialPsychology:1987}, or a single simulated courtroom trial \cite{Erickson+al:78a}.
A recent large-scale study of language coordination in the on-line domain
\citep{Danescu-Niculescu-Mizil+al:11a} used data from Twitter, 
where markers of status and power are not as readily inferred;
they identified a weak correlation between language coordination
and Twitter follower counts, suggesting a potential 
connection to status measures.  Additionally, researchers have used text features other than
linguistic coordination to identify status differences
\citep{Diehl+Namata+Getoor:2007a,McCallum+Wang+CorradaEmmanuel:2007a,scholand2010social,Bramsen+al:2011a,Gilbert:ProceedingsOfCscw:2012}; in contrast with our work, these methods picked up situation-specific cues, such as the word ``termination'' for the
Enron corporate-email corpus \citep{Diehl+Namata+Getoor:2007a}, which are
unlikely to generalize across contexts.  

\vpara{Collaborative communities} 
Interaction in online communities has been extensively studied.
Wikipedia was used as a testbed for studying user interaction at large \cite{billings2010understanding,Lu:2010:ESC:1772690.1772761,Viegas+al:2007a,taraborellibeyond,Laniado+al:2011a} and the promotion process in such communities \cite{Burke+Kraut:2008a,Leskovec+Huttenlocher+Kleinberg:2010a}. Reviewer behavior and incentives to participate in the collaborative process were studied in the context of commercial review sites \cite{Gilbert2009,wu2010opinion,Bryant:2005:BWT:1099203.1099205,Lu:2010:ESC:1772690.1772761}.


\small
\xhdr{Acknowledgments}
We thank Timothy Hawes for sharing the processed Supreme Court 
transcripts from \cite{Hawes:JournalOfTheAmericanSocietyForInformation:2009},
and 
\newcommand{\fn}[2]{#1#2}
\fn{B}{runo} Abrahao, 
\fn{E}{ric} Baumer, 
\fn{C}{laire} Cardie,
\fn{E}{unsol} Choi, 
\fn{C}{hris} Diehl, 
\fn{S}{usan} Dumais,
\fn{S}{himon} Edelman, 
\fn{J}{acob} Eisenstein,
\fn{M}{ichael} Gamon,
\fn{S}{usan} Herring, 
\fn{M}{olly} Ireland, 
\mbox{\fn{D}{iana}} Minculescu, 
\fn{A}{lex} Niculescu-Mizil,
\fn{M}{yle} Ott,
\fn{J}{on} Park,
\fn{P}{hilip} Resnik,
\fn{D}{aniel} Romero,
\fn{C}{henhao} Tan,
\fn{L}{u} Wang, 
\fn{B}{ishan} Yang, 
\fn{A}{inur} Yessenalina, 
and the anonymous reviewers
 for valuable discussions and suggestions.
 This paper is based upon work supported in part by the NSF grant
 IIS-0910664, 
IIS-1016099, and grants from Google and Yahoo!.

 \bibliographystyle{abbrv-shrink}
\renewcommand*{\bibfont}{\raggedright}
\newcommand{\bibsnip}{\vspace*{-.04in}}

\end{document}